\documentclass[intlimits,twoside,a4paper]{article}

\usepackage{amsmath,amssymb}
\usepackage{graphicx}

\usepackage[T2A]{fontenc}
\usepackage[cp1251]{inputenc}

\usepackage[eqsecnum]{cmpj2}

\issue{2013}{16}{1}{13701}

\doinumber{10.5488/CMP.16.13701}

\title[Electronic properties of AlN crystal doped with Cr, Mn and Fe]%
{Electronic properties of AlN crystal \\ doped with Cr, Mn and Fe}
\author[S.V.~Syrotyuk, V.M.~Shved]{S.V.~Syrotyuk, V.M.~Shved}
\address{Lviv Polytechnic National University, 12~S.~Bandera St., 79013~Lviv, Ukraine}

\date{Received  May 24, 2012, in final form August 20, 2012}
\authorcopyright{S.V.~Syrotyuk, V.M.~Shved, 2013}

\begin{document}

\maketitle

\begin{abstract}

The spin-resolved electronic energy band spectra, as well as partial and total
density of electronic states of the crystal AlN, doped with Cr, Mn and Fe,
have been evaluated within the projector augmented waves (PAW) approach by means of the ABINIT code. The Hartree-Fock exchange for correlated electrons is used to describe the correlated orbitals in the PAW framework. The calculated one-electron energies for electrons of spin up and down are very different. We have found that all the considered crystals are ferromagnetic.

\pacs 71.55.Eq, 71.55.$-$i, 71.20.$-$b, 71.15.Mb, 71.27.$+$a, 75.30.$\pm$m
\end{abstract}

\section{Introduction}

Semiconductors doped with $d$-transition elements draw considerable attention of researchers due to their possible applications in spintronics~\cite{b1}. The discovery of ferromagnetism therein led to an intensification of experimental and theoretical studies. The impurities of transition elements lead to a fundamental change in the electronic structure of semiconductors~\cite{b2}. Crystal AlN with an admixture of 3$d$-transition elements was investigated using the ultrasoft (USPP)  pseudopotential approach within the LSDA theory~\cite{b3} and in the LSDA${}+U$ approximation~\cite{b4}. The partial and total densities of electronic states of the Mn-doped crystal AlN and co-doped with B, Si, and P, with spin up and down~\cite{b5}, have been recently evaluated by means of the LSDA KKR program~\cite{b6}. The parameter of the Coulomb energy $U$ makes it possible to take into account strong correlations between the $d$-electrons of the transition element, and thus to improve the values of the electronic energy band spectrum of the crystal with impurity.
	
However, it is known that the Coulomb energy $U$ depends on the system. This means that it cannot be used in  calculations in the same crystal, for example, with vacancies~\cite{b7}. As it is inherent to LSDA${}+U$ approach, in~\cite{b7} there was also found the existence of several minima of the total energy functional depending on the matrix of the initial occupation of the correlated orbitals.
	
The spin-resolved electronic energy band spectra, as well as partial and total density of electronic states of the crystal AlN, doped with Cr, Mn and Fe, have been evaluated within the projector augmented wave (PAW) approach~\cite{b8,b9}, implemented in the ABINIT~\cite{b10} code. The exact exchange, for 3$d$-electrons only, is used to improve the on-site correlations~\cite{b11}.

\section{Calculation}

% \subsection{Title information}

The electronic structure calculations have been carried out using the PAW~\cite{b9} method, in which the electron wave function with its full nodal structure $|\psi_{n}(\mathbf{r})\rangle$ is represented in terms of the smooth nodeless function $|\tilde{\psi}_{n}(\mathbf{r})\rangle$ as follows:
\begin{equation}\label{1}
|\psi_{n}(\mathbf{r})\rangle = \tau |\tilde{\psi}_{n}(\mathbf{r})\rangle,
\end{equation}
where $\tau$ is the transformation operator,
\begin{equation}\label{2}
\tau = 1+\sum_a \sum_\alpha \left(|\phi_{\alpha}^{a}\rangle-|\tilde\phi_{\alpha}^{a}\rangle\right)\langle\tilde {p}_{\alpha}^{a}|\,,
\end{equation}
and $|\phi_{\alpha}^{a}(\mathbf{r})\rangle$ is the all-electron basis function, $|\tilde\phi_{\alpha}^{a}(\mathbf{r})\rangle$ is the pseudopotential basis function, and $|\tilde{p}_{\alpha}^{a}(\mathbf{r})\rangle$ is projector function, which we calculated using the \emph{atompaw} code~\cite{b12}. The subscript $\alpha$ denotes the quantum numbers, and the superscript $a$ represents the atomic augmentation sphere index.

We have generated the PAW functions for the following valence basis states: ${2s^22p^63s^23p^1}$ for Al, ${2s^22p^3}$ for N, ${3s^23p^63d^54s^14p^0}$ for Cr, ${3s^23p^63d^54s^24p^0}$ for Mn, and ${3s^23p^63d^64s^24p^0}$ for Fe. The inclusion in the basis of 3$s$ and 3$p$ states of the Cr, Mn and Fe atoms slows down the calculation, but increases the precision and eliminates the problem of ghost states which can appear among the solutions of a secular equation constructed on a minimal set of functions~\cite{b13}. The radii of the augmentation spheres are 1.6, 1.3, 1.9, 1.9, 1.9 a.u. for Al, N, Cr, Mn, Fe, respectively.
Substituting (\ref{1}), (\ref{2}) to the Kohn-Sham equation
\begin{equation}\label{3}
\emph{H}|\psi_{\alpha\mathbf{k}}\rangle = |\psi_{\alpha\mathbf{k}}\rangle\varepsilon_{\alpha\mathbf{k}}
\end{equation}
we obtain the system of linear equations
\begin{equation}\label{4}
\tau^+\emph{H}\tau|\tilde\psi_{\alpha\mathbf{k}}\rangle = \tau^+\tau|\tilde\psi_{\alpha\mathbf{k}}\rangle\varepsilon_{\alpha\mathbf{k}}\,,
\end{equation}
where $\alpha$ is a band number, $\mathbf{k}$ denotes the vector in the first Brillouin zone, and $\tau^+$ defines Hermitian conjugate of $\tau$.
The electron density in the PAW method is determined by the sum of three terms,
\begin{equation}\label{5}
\rho(\mathbf{r}) = \tilde\rho(\mathbf{r})+\sum_a\left[\rho^a(\mathbf{r})-\tilde\rho^a(\mathbf{r})\right].
\end{equation}
The first term is a smooth pseudo-density, which is the Fourier series,
\begin{equation}\label{6}
\tilde\rho(\mathbf{r}) = \sum_{n,\mathbf{k}}f_{n\mathbf{k}}|\tilde\psi_{\alpha\mathbf{k}}(\mathbf{r})|^2 = \frac{1}{\Omega}\sum_\mathbf{G}\tilde\rho(\mathbf{G})\re^{\ri\mathbf{G}\mathbf{r}}\, ,
\end{equation}
where $f_{n\mathbf{k}}$ is occupancy, weighted by the fractional Brillouin zone sampling volume. The terms of the density of electrons in the atomic spheres are determined from the coefficients of the projected population of the states
\begin{equation}\label{7}
W_{\alpha\beta}^{a}=\sum_{n,\mathbf{k}}f_{n\mathbf{k}}
\langle{\tilde\psi_{\alpha\mathbf{k}}|\tilde{p}_\alpha^a}\rangle\langle{\tilde{p}_\beta^a}|\tilde\psi_{\alpha\mathbf{k}}\rangle.
\end{equation}
Hence, the contributions of the density in the atomic sphere can be written as
\begin{equation}\label{8}
\rho^a(\mathbf{r})=\sum_{\alpha\beta}W_{\alpha\beta}^a\varphi_\alpha^{a^*}(\mathbf{r})
\varphi_\beta^{a}(\mathbf{r}), \qquad  \tilde\rho^a(\mathbf{r})=\sum_{\alpha\beta}W_{\alpha\beta}^a\tilde\varphi_\alpha^{a^*}(\mathbf{r})
\tilde\varphi_\beta^{a}(\mathbf{r}).
\end{equation}
Exchange-correlation potential was calculated in the form of PBE0~\cite{b14} according to which the exchange-correlation energy
\begin{equation}\label{9}
E_\mathrm{xc}^\mathrm{PBE0}[\rho]=E_\mathrm{xc}^\mathrm{PBE}[\rho]+\frac{1}{4}\left(E_\mathrm{x}^\mathrm{HF}
[\psi_\mathrm{sel}]-E_\mathrm{x}^\mathrm{PBE}[\rho_\mathrm{sel}]\right),
\end{equation}
corresponds to the PBE exchange-correlation functional~\cite{b15}, and $\psi_\mathrm{sel}$, $\rho_\mathrm{sel}$ represent the wave function and electron density of the selected electrons, respectively~\cite{b16}. The latter are the 3$d$-electrons of Cr, Mn and Fe.

The electronic energy bands and DOS have been evaluated by means of the ABINIT code~\cite{b10}. Integration over the Brillouin zone was performed on the Monkhorst-Pack~\cite{b17} spatial grid of $4\times6\times6$. The lattice constant of the crystal AlN $a = 4.36$~{\AA}, and the supercell parameters containing 16 atoms are $2\times1\times1$. The iterations were performed to ensure the calculation of the total energy of the crystal with an accuracy of $10^{-8}$~Ha. Relativistic effects are taken into account within the scalar relativistic approximation. The density of electronic states was evaluated from the equation
\begin{equation}\label{10}
n_\beta(E)=\sum_{\alpha,\mathbf{k}}\delta(E-E_{\alpha\mathbf{k}})|P_{\alpha\beta}^a(\mathbf{k})|^2,
\end{equation}
\begin{equation}\label{11}
P_{\alpha\beta}^a(\mathbf{k})=\langle{\tilde{p}_\beta^a}|\tilde\psi_{\alpha\mathbf{k}}\rangle
=\int{\rd}\mathbf{r}\tilde{p}_\beta^{a^*}\left(\mathbf{r}-\mathbf{R}^a\right)\tilde\psi_{\alpha\mathbf{k}}(\mathbf{r}),
\end{equation}
where $P_{\alpha\beta}^a(\mathbf{k})$ are the expansion coefficients  for both
\begin{equation}\label{12}
|\tilde\psi_{\alpha\mathbf{k}}\rangle = \sum_\beta{P_{\alpha\beta}^a(\mathbf{k})}|\tilde\phi_\beta^a\rangle,
\end{equation}
\begin{equation}\label{13}
|\psi_{\alpha\mathbf{k}}\rangle = \sum_\beta{P_{\alpha\beta}^a(\mathbf{k})}|\phi_\beta^a\rangle,
\end{equation}
smooth and true functions, respectively, within the augmentation sphere. As can be seen from equation~(\ref{9}), the exchange-correlation energy functional depends on the density of electrons and on the selected wave function of strongly correlated electrons. First, the pseudo-wave function is calculated from equations~(\ref{4}). Then, from equation~(\ref{1}), the electron wave function is calculated and the density of electrons~(\ref{5}) is obtained. Based on the latter we find the exchange-correlation energy functional~(\ref{9}). The symmetry of the crystal is described by space group P-42m (number 111) and the Bravais lattice is tP (primitive tetragonal).

%%%%%%%%%%%%%%%%%%%%%%%%%%%

\begin{figure}[!b]
\centerline{
\includegraphics[width=0.45\textwidth]{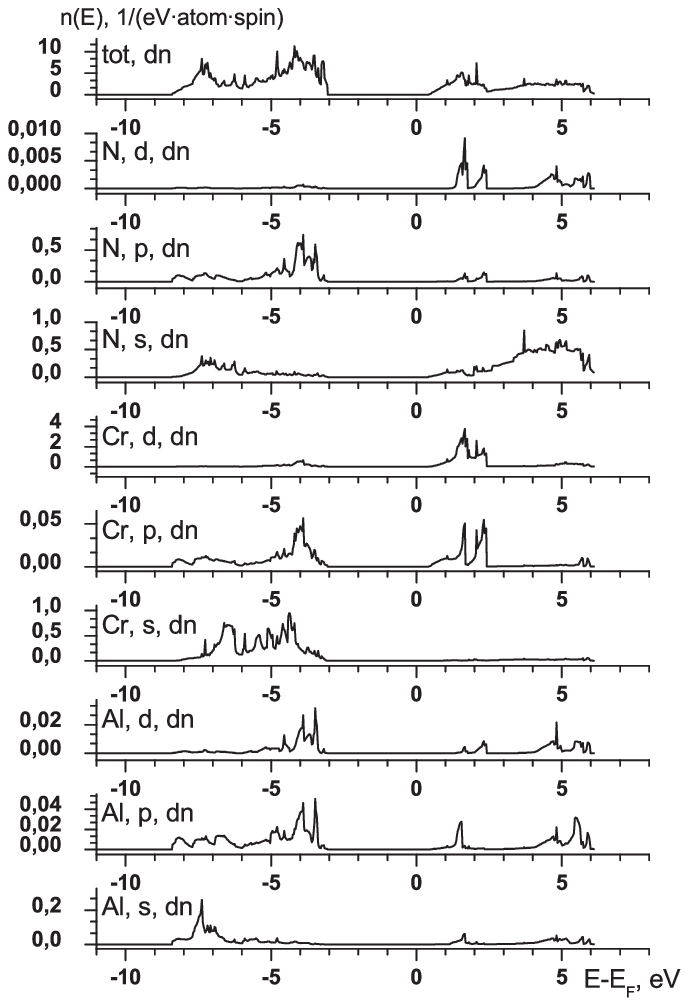}
\hspace{8mm}%\hfill%
\includegraphics[width=0.45\textwidth]{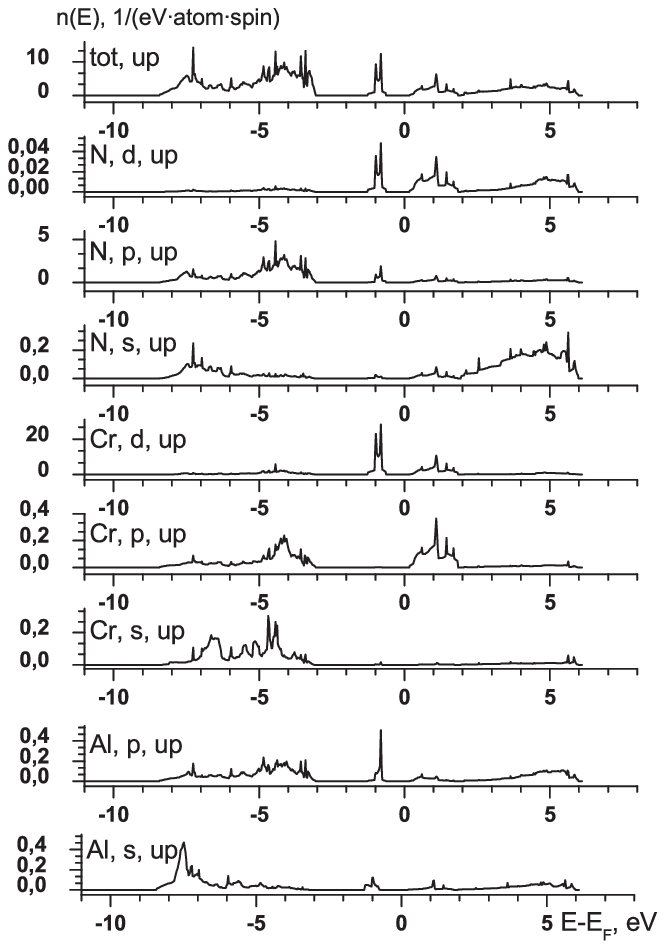}
}
\vspace{-3mm}
\parbox[t]{0.50\textwidth}{  \caption{The spin down partial, in 1/(eV·atom·spin), and total densities of states for Al$_7$Cr$_1$N$_8$.}\label{f1}}
\parbox[t]{0.50\textwidth}{  \caption{The spin up partial, in 1/(eV·atom·spin), and total densities of states for Al$_7$Cr$_1$N$_8$.}\label{f2}
}
\end{figure}

%%%%%%%%%%%%%%%%%%%%%%

\section{Electronic properties}

\begin{figure}[!t]
\centerline{
\includegraphics[width=0.41\textwidth]{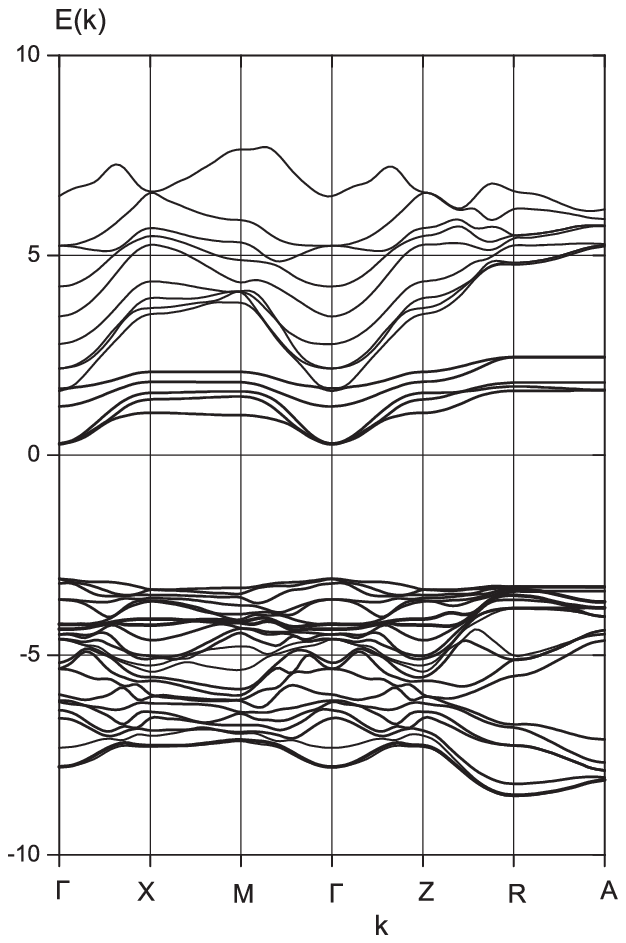}
\hspace{12mm}%\hfill%
\includegraphics[width=0.41\textwidth]{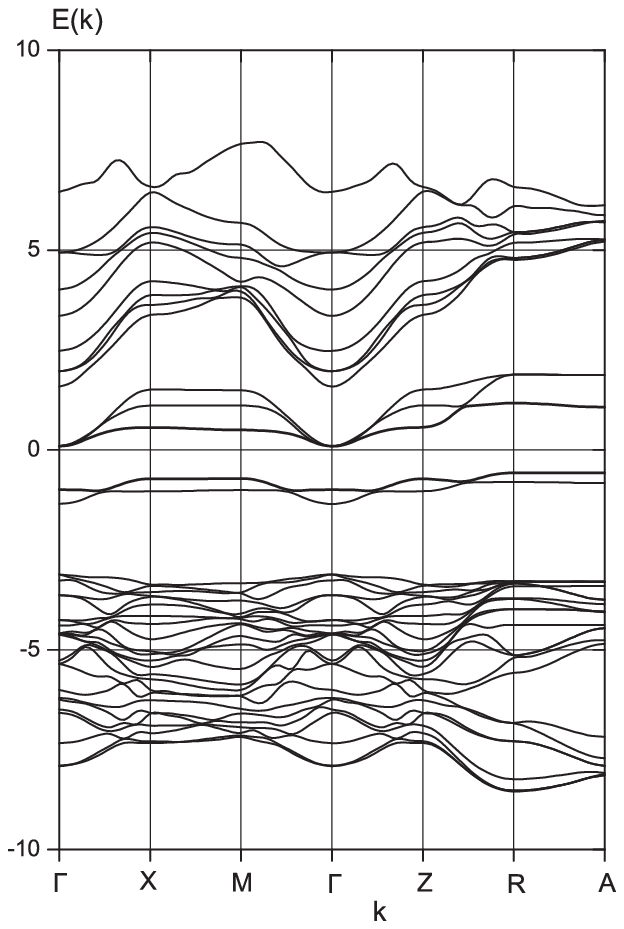}
}
\vspace{-2mm}
\parbox[t]{0.50\textwidth}{
  \caption{The spin down band structure of Al$_7$Cr$_1$N$_8$.}\label{f3}
  }
\parbox[t]{0.50\textwidth}{
  \caption{The spin up band structure of Al$_7$Cr$_1$N$_8$.}\label{f4}
}
\end{figure}

Figure~\ref{f1} shows the partial and total density of electronic states of the crystal Al$_7$Cr$_1$N$_8$ for a spin down. We see that the top of the valence band is formed predominantly with hybridized $s$-states of Cr and $p$-states of N. The dispersion curves shown in figure~\ref{f3}, indicate that for spin down, our crystal is a semiconductor with a direct gap. It is substantially less than the gap $\Gamma$-X in the usual AlN crystal. Aluminum nitride in the zinc-blende structure is an insulator with a wide indirect energy bandgap of 5.3~eV. Such a narrowing is explained by the fact that the crystal CrN is a Mott-Hubbard-type insulator with a small to negligible indirect band gap~\cite{b18}. The bottom of the valence band is now at the point R, while in the the usual crystal AlN, it is at the point $\Gamma$. The Fermi level is located in the band gap, quite close to the bottom of the conduction band.

For spin up, we have a completely different picture, represented in figures~\ref{f2}, and~\ref{f4}. The crystal Al$_7$Cr$_1$N$_8$ is a semiconductor with indirect narrow gaps $\Gamma$-R and $\Gamma$-A. The Fermi level is now even closer to the bottom of the conduction band. The top of the valence band is formed by hybridized $d$-states of Cr, $p$-states of N and Al.

\begin{figure}[!t]
%\vspace{1mm}
\centerline{
\includegraphics[width=0.445\textwidth]{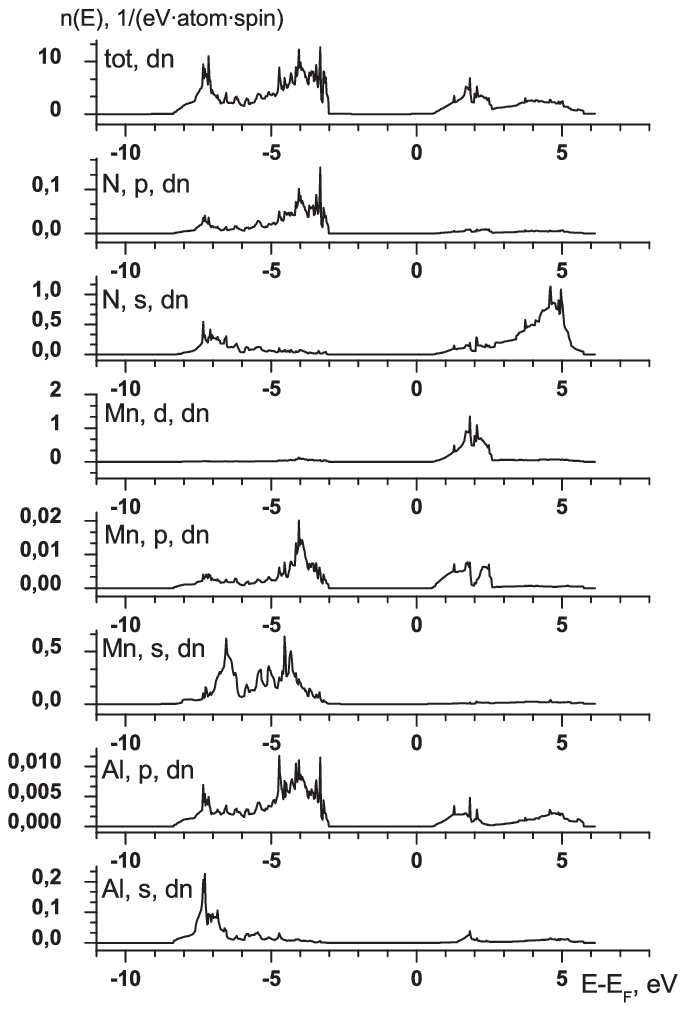}
\hspace{8mm}%\hfill%
\includegraphics[width=0.44\textwidth]{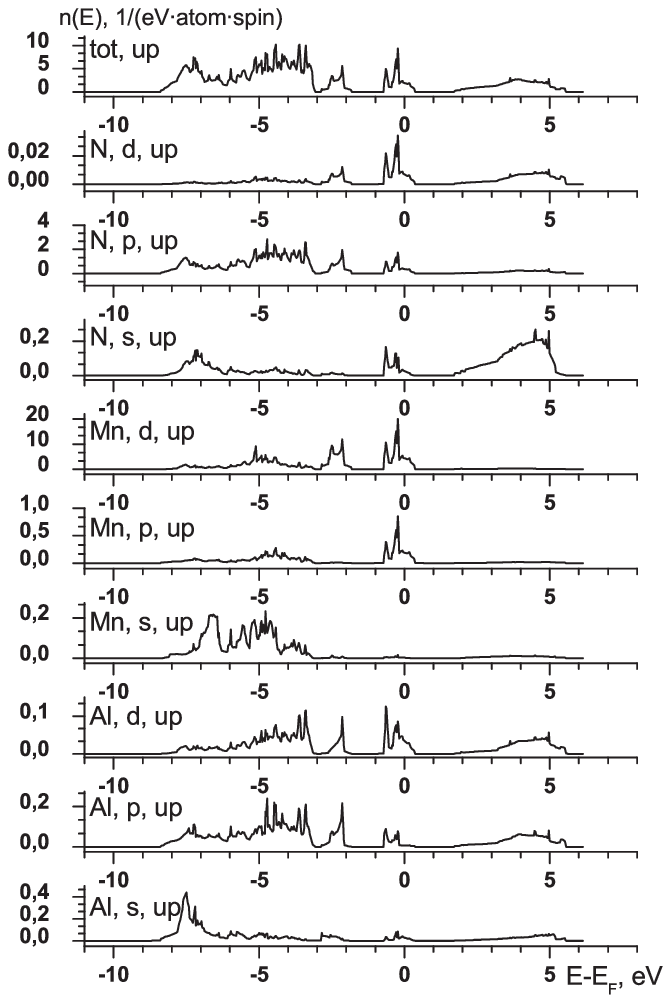}
}
\vspace{-4mm}
\parbox[t]{0.50\textwidth}{
  \caption{The spin down partial, in 1/(eV·atom·spin), and total densities of states for Al$_7$Mn$_1$N$_8$.}\label{f5}
  }
\parbox[t]{0.50\textwidth}{
  \caption{The spin up partial, in 1/(eV·atom·spin), and total densities of states for Al$_7$Mn$_1$N$_8$.}\label{f6}
  }
\end{figure}

\begin{figure}[!b]
\centerline{
\includegraphics[width=0.41\textwidth]{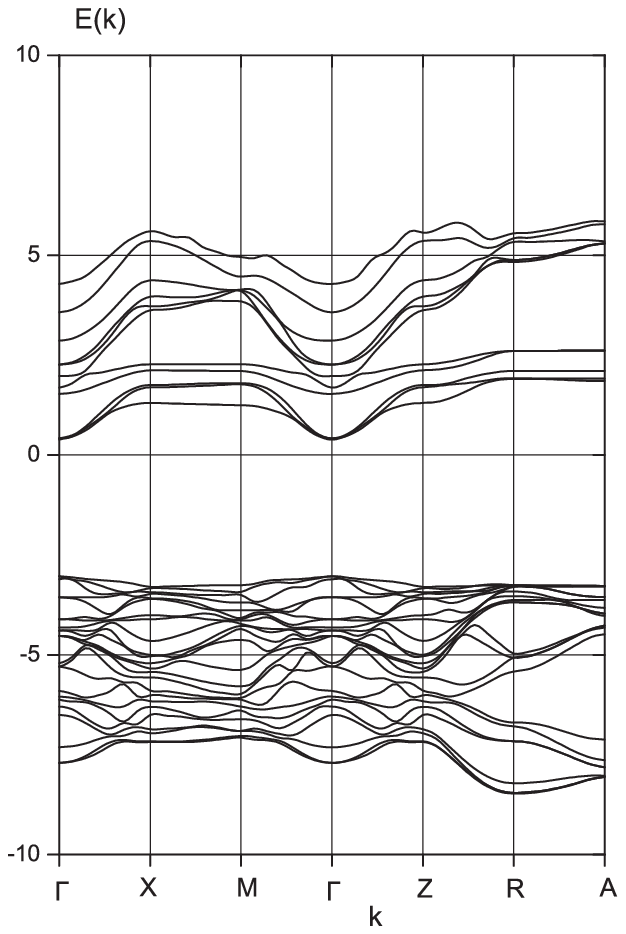}
\hspace{10mm}%
\includegraphics[width=0.41\textwidth]{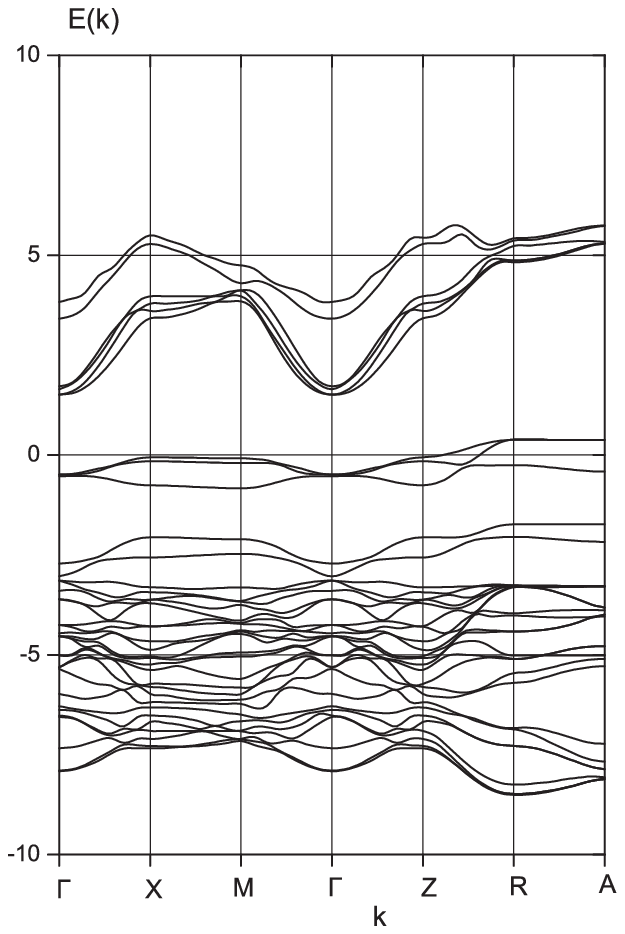}
}
\vspace{-4mm}
\parbox[t]{0.50\textwidth}{
  \caption{The spin down band structure of Al$_7$Mn$_1$N$_8$.}\label{f7}
  }
\parbox[t]{0.50\textwidth}{
  \caption{The spin up band structure of Al$_7$Mn$_1$N$_8$.}\label{f8}
  }
\end{figure}

We turn to the analysis of the results of calculation for the Al$_7$Mn$_1$N$_8$ crystal, represented in fig-\linebreak ures~\ref{f5}--\ref{f8}.

Figures~\ref{f5} and~\ref{f7} describe the electronic density of states and energy bands for spin down. From figure~\ref{f5} we see that the top of the valence band forms the $s$-states of Mn and the $p$-states of N. Figure~\ref{f7} shows that the crystal is a direct band gap semiconductor, with the bottom of the valence band at the point R.

Figures~\ref{f6} and~\ref{f8} describe the electronic density of states and energy bands for spin up. Figure~\ref{f6} shows that the top of the valence band is formed by the $p$- and $d$-states of Mn and the $p$-states of N, but the bottom of the conduction band forms the $s$-states of N. Figure~\ref{f8} shows that the crystal is a semiconductor with indirect gap $\Gamma$-ZR. The Fermi level is at the top of the valence band. The presence of the spin up 3$d$ states of Mn at Fermi level (figure~\ref{f8}) and of the spin down ones in the conduction band (figure~\ref{f6}) qualitatively agrees with the results obtained within the DFT${}+U$ approach for Mn doped AlN~\cite{b4}.


%%%%%%%%%%%%%%%%%%%%%%%%%%%%%%%%%%%%%%%%%%%%%

\begin{figure}[!t]
\centerline{
\includegraphics[width=0.42\textwidth]{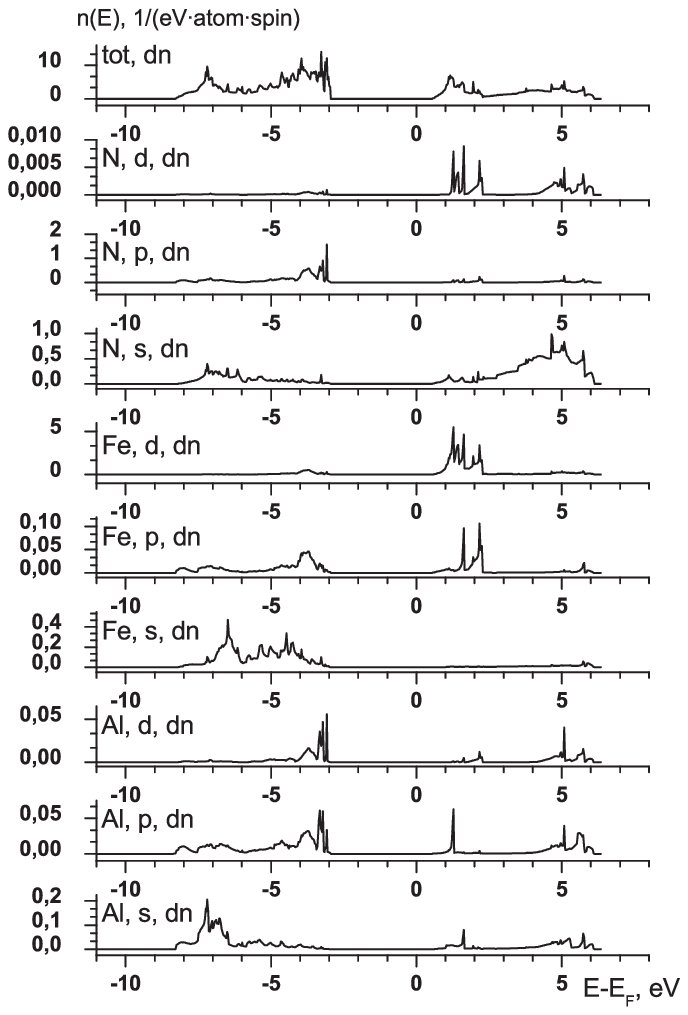}
\hspace{10mm}%\hfill%
\includegraphics[width=0.42\textwidth]{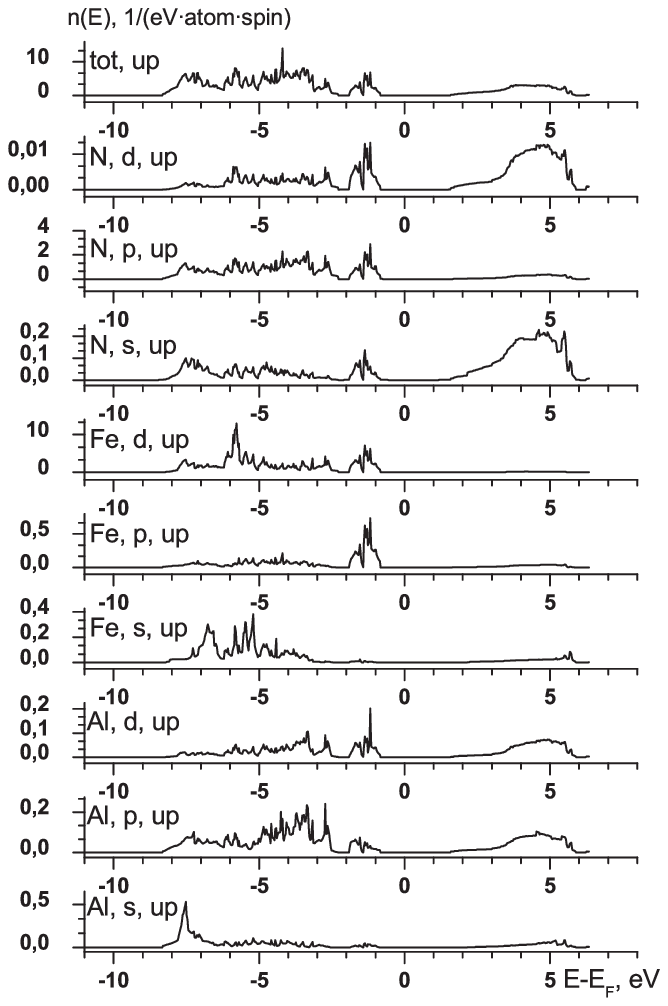}
}
\vspace{-2mm}
\parbox[t]{0.50\textwidth}{
  \caption{The spin down partial, in 1/(eV·atom·spin), and total densities of states for Al$_7$Fe$_1$N$_8$.}\label{f9}
  }
\parbox[t]{0.50\textwidth}{
  \caption{The spin up partial, in 1/(eV·atom·spin), and total densities of states for Al$_7$Fe$_1$N$_8$.}\label{f10}
  }
\end{figure}
\begin{figure}[!b]
\centerline{
\includegraphics[width=0.41\textwidth]{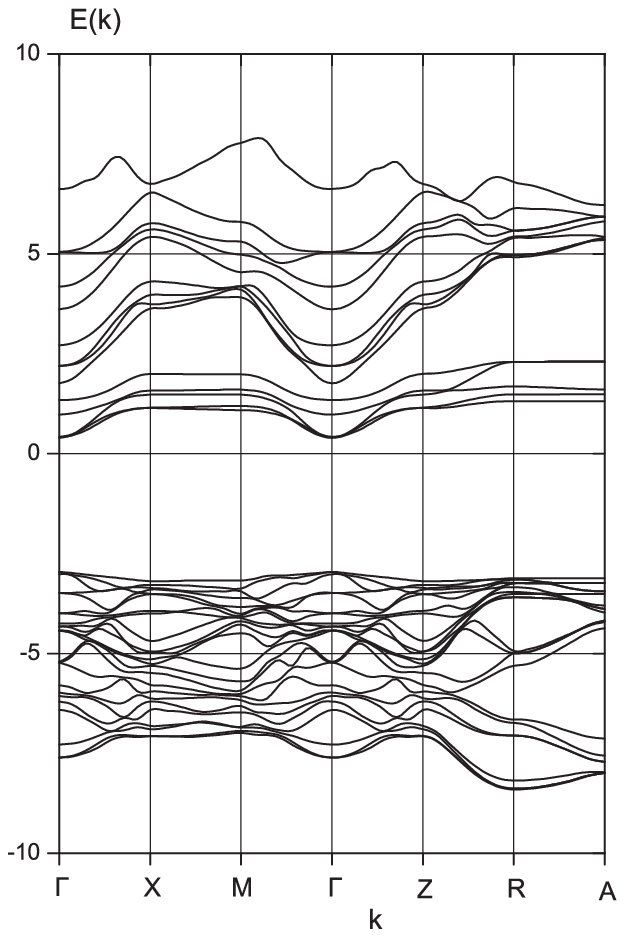}
\hspace{10mm}%\hfill%
\includegraphics[width=0.41\textwidth]{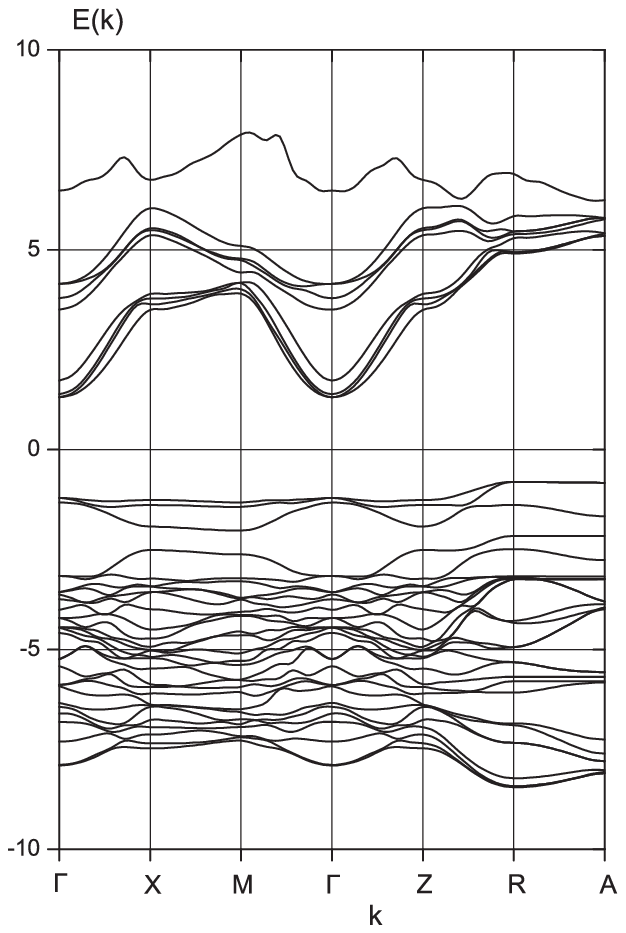}
}
\vspace{-2mm}
\parbox[t]{0.50\textwidth}{
  \caption{The spin down band structure of Al$_7$Fe$_1$N$_8$.}\label{f11}
}
\parbox[t]{0.50\textwidth}{
  \caption{The spin up band structure of Al$_7$Fe$_1$N$_8$.}\label{f12}
}
\end{figure}

Let us consider the results of the calculation for the Al$_7$Fe$_1$N$_8$ crystal, presented in figures~\ref{f9}--~\ref{f12}. The calculation results for the spin down are presented in figures~\ref{f9} and~\ref{f11}. Figure~\ref{f9} shows that the top of the valence band is formed by $p$-states of N, and the bottom of the conduction band consists predominantly of $d$-states of Fe. Figure~\ref{f11} shows the direct band gap at the $\Gamma$ point. The Fermi level is located near the bottom of the conduction band. The bottom of the valence band is at the point R.

The calculation results for the spin up are presented in figures~\ref{f10} and~\ref{f12}. In figure~\ref{f10} it is seen that the top of the valence band is formed by $d$-states of Fe and $p$-states of N. The bottom of the conduction band forms the $p$-states of Al and $s$-states of N. From figure~\ref{f12} we see that the crystal is a semiconductor with indirect gaps $\Gamma$-R and $\Gamma$-A. The bottom of the valence band is at the point~R.

\section{Conclusions}

For the spin down, all the crystals are direct band gap semiconductors, with direct gap at the point $\Gamma$. And for the spin up all the crystals are characterized by indirect band gap. All the crystals show the ferromagnetic ordering. For AlCrN, our result is confirmed by experimental data on ferromagnetism detected above the room temperature \cite{b19}.  For AlMnN we observe a half-metallic behavior in the sense that the Fermi level state density is finite for the majority spin, and zero for the minority spin, as it is seen in figures~\ref{f7} and~\ref{f8}. Figure~\ref{f6} shows that at the Fermi level, there are the spin-polarized carriers.  This is well matched with the results obtained in the work~\cite{b20} within the LSDA approach. However, our calculation shows that the AlCrN crystal does not possess the half-metallicity predicted in the work~\cite{b20}. We have calculated the electron energy spectrum and the density of electronic states in the PAW GGA formalism without accounting the strong correlations of 3$d$-electrons. It was found that there is no state of half-metallicity in the AlCrN crystal which does exist in the crystals of AlMnN and AlFeN. Meanwhile, for the latter crystal, figures~\ref{f11} and~\ref{f12}, obtained within the PAW PBE0 approach, the half-metallic state is not observed. As we see in table~\ref{tbl}, the values of both direct and indirect gaps of the crystals are quite different. The largest contributions to the density of states at the Fermi level is provided by  the states of $d$- and $p$-symmetry of manganese and $p$-states of nitrogen. The difference between the results obtained herein with the account of strong electron correlations, in the PBE0 approach, and in LSDA theory, is due to inadequacy of the latter in the description of the wave functions and energy levels of 3$d$-electrons.

\begin{table}[htb]
\caption{The calculated electronic band gaps (for spin down, dn, for spin up, up, in eV) and spin magnetic moments (\emph{m} in $\mu_{\rm B}$) of crystals.}
\label{tbl}
\vspace{2ex}
\begin{center}
\renewcommand{\arraystretch}{0}
\begin{tabular}{|c||c|c|c|}
\hline\hline
& Al$_7$Cr$_1$N$_8$ & Al$_7$Mn$_1$N$_8$ & Al$_7$Fe$_1$N$_8$\strut\\
\hline
\rule{0pt}{2pt}&&&\\
\hline
 $E_{\rm c}-E_{\rm v}, \rm{dn}$ & 3.36, direct & 3.42, direct & 3.35, direct\strut\\
\hline
 $E_{\rm c}-E_{\rm v}, \rm{up}$ & 0.66, indirect & 1.51, indirect & 2.13, indirect\strut\\
\hline
 $E_{\rm c}-E_{\rm F}, \rm{dn}$& 0.27 & 0.38 & 0.39\strut\\
\hline
 $E_{\rm c}-E_{\rm F}, \rm{up}$ & 0.09 & 1.51 & 1.30\strut\\
\hline
 $E_{\rm F}-E_{\rm v}, \rm{dn}$ & 3.09 & 3.04 & 2.96\strut\\
\hline
 $E_{\rm F}-E_{\rm v}, \rm{up}$ & 0.57 & 0.00 & 0.83\strut\\
\hline
 $\emph{m}, \rm{cell}$ & 1.3 & 2.7 & 3.3\strut\\
\hline
 $\emph{m}, \rm{local}$ & 1.1 (Cr) & 2.4 (Mn) & 3.3 (Fe)\strut\\
\hline\hline
\end{tabular}
\renewcommand{\arraystretch}{1}
\end{center}
\end{table}

As a result, the calculation from first principles of electronic and magnetic properties of the AlTN systems, where T is a transition element has been made taking into account strong correlations of 3$d$-electrons. A significant narrowing of the interband gap was found for all systems. The calculated electron energy spectrum and the density of electronic states, as well as the value of the Fermi energy are important parameters for the systems considered to be essential candidates for high-temperature electronics and optoelectronic applications~\cite{b21}.

Another important factor is the issue of maintaining  the ferromagnetic state at  higher temperatures as much as possible. However, this is the subject of a separate study.

%\clearpage

\ukrainianpart

\title{Електронні властивості кристала AlN, \\ легованого Cr, Mn та Fe}
\author{С.В. Сиротюк, В.М. Швед}
\address{Національний університет "Львівська політехніка", вул. С. Бандери 12, 79013 м. Львів, Україна}
%
%% якщо автор є один або автори є з однієї установи:
%
%  \author{1й Автор, 2й Автор, \ldots}
%  \address{Інститут\ldots}
%
\makeukrtitle

\begin{abstract}
\tolerance=3000%
Залежні від спіна електронні енергетичної спектри,
а також парціальні й повні щільності електронних станів кристала AlN,
легованого Cr, Mn та Fe, були отримані за методом проекційних парціальних хвиль (PAW)
за допомогою програми ABINIT. Обмінний потенціал Хартрі-Фока для скорельованих електронів
використовується для опису скорельованих орбіталей в рамках PAW.
Розраховані одноелектронні енергії для електронів зі спіном вгору і вниз є дуже різні.
Ми виявили, що всі розглянуті кристали є феромагнетиками.
\keywords розрахунок електронної структури, сильно скорельовані електрони, точний обмін для скорельованих електронів, магнітні напівпровідники, метод проекційних парціальних хвиль

\end{abstract}


\begin{thebibliography}{99}

\bibitem{b1} Newman N., Wu S.Y., Liu H.X., Medvedeva J., Gu L., Singh R.K., Yu Z.G., Krainsky I.L., Krishnamurthy S., Smith~D.J., Freeman A.J., van Schilfgaarde M., Phys. Status Solidi A, 2006, \textbf{203}, 2729; \doi{10.1002/pssa.200669636}.

\bibitem{b2} Tao Zhi-Kuo, Zhang Rong, Cui Xu-Gao, Xiu Xiang-Qian, Zhang Guo-Yu, Xie Zi-Li, Gu Shu-Lin, Shi Yi, Zheng You-Dou, Chin. Phys. Lett., 2008, \textbf{25}, 1476; \doi{10.1088/0256-307X/25/4/084}.

\bibitem{b3}Kaczkowski J., Jezierski A., Acta Phys. Pol. A, 2009, \textbf{115}, 275.

\bibitem{b4}Kaczkowski J., Jezierski A., Acta Phys. Pol. A, 2009, \textbf{116}, 924.

\bibitem{b5}Syrotyuk S.V., Shved V.M., Ukr. J. Phys., 2011, \textbf{56}, 564.

\bibitem{b6}Akai H., Phys. Rev. Lett., 1998, \textbf{81}, 3002; \doi{10.1103/PhysRevLett.81.3002}.


\bibitem{b7}Jollet F., Jomard G., Amadon B., Crocombette J.P., Torumba D., Phys. Rev. B, 2009, \textbf{80}, 235109; \\ \doi{10.1103/PhysRevB.80.235109}.

\bibitem{b8}Bl\"{o}chl P.E., Phys. Rev. B, 1994, \textbf{50}, 17953; \doi{10.1103/PhysRevB.50.17953}.

\bibitem{b9}Tackett A.R., Holzwarth N.A.W., Matthews G.E., Comput. Phys. Commun., 2001, \textbf{135},  348; \\ \doi{10.1016/S0010-4655(00)00241-1}.

\bibitem{b10}Gonze X. %, Amadon B., Anglade P.-M., Beuken J.-M.
et al., Comput. Phys. Commun., 2009, \textbf{180}, 258; \doi{10.1016/j.cpc.2009.07.007}.

\bibitem{b11}Nov\'{a}k P., Kune\v{s} J., Chaput L., Pickett W.E., Phys. Status Solidi B, 2006, \textbf{243}, 563; \doi{10.1002/pssb.200541371}.

\bibitem{b12}Holzwarth N.A.W., Tackett A.R., Matthews G.E., Comput. Phys. Commun., 2001, \textbf{135},  329; \\ \doi{10.1016/S0010-4655(00)00244-7}.

\bibitem{b13}Holzwarth N.A.W., Matthews G.E., Dunning R.B., Tackett A.R., Zeng Y., Phys. Rev. B, 1997, \textbf{55}, 2005; \\  \doi{10.1103/PhysRevB.55.2005}.

\bibitem{b14}Ernzerhof M., Scuseria G.E., J. Chem. Phys., 1999, \textbf{110}, 5029; \doi{10.1063/1.478401}.

\bibitem{b15}Perdew J.P., Burke K., Ernzerhof M., Phys. Rev. Letters, 1996, \textbf{77}, 3865; \doi{10.1103/PhysRevLett.77.3865}.

\bibitem{b16}Tran E., Blaha P., Schwarz K., Novak P., Phys. Rev. B, 2006, \textbf{74}, 155108; \doi{10.1103/PhysRevB.74.155108}.

\bibitem{b17}Monkhorst H.J., Pack J.D., Phys. Rev. B, 1976, \textbf{13}, 5188; \doi{10.1103/PhysRevB.13.5188}.

\bibitem{b18}Zhang X.Y., Gall D., Phys. Rev. B, 2010, \textbf{82}, 045116; \doi{10.1103/PhysRevB.82.045116}.

\bibitem{b19}Song Y.Y., Quang P.H., Lim K.S., Yu S.C., J. Korean Phys. Soc., 2006, \textbf{48}, 1449.

\bibitem{b20}Chen H., Zhang J.-F., Yuan H.-K., Commun. Theor. Phys., 2007, \textbf{48}, 749; \doi{10.1088/0253-6102/48/4/037}.

\bibitem{b21}Zhao L., Lu P.-F., Yu Z.-Y., Guo X.-T., Ye H., Yuah G.-F., Shen Y., Liu Y.-M., Commun. Theor. Phys., 2011, \textbf{55}, 893; \doi{10.1088/0253-6102/55/5/29}.

\end{thebibliography}
\end{document}